\documentclass[twocolumn,aps,prc,superscriptaddress,showpacs,nofootinbib,floatfix]{revtex4}

\usepackage{graphicx}		

\newcommand{\pt}{p_T}
\newcommand{\dau}{$d+$Au}

\newcommand{\pa}{$p+$A}
\newcommand{\myaa}{{\rm A}+{\rm A}}

\newcommand{\auau}{Au$+$Au}
\newcommand{\pbpb}{Pb$+$Pb}

\newcommand{\pp}{p+p}

\newcommand{\jpsi}{J/\psi}

\newcommand{\npart}{N_{\rm part}}
\newcommand{\ncoll}{N_{\rm coll}}
\newcommand{\rdau}{R_d{\rm Au}}
\newcommand{\raa}{R_{\rm AA}}
\newcommand{\sqrts}{\sqrt{s_{_{NN}}}}

\newcommand{\ccbar}{$c\overline{c}$}

\begin{document}
 
\title{$J/\psi$ suppression at forward rapidity in Au$+$Au collisions at 
       $\sqrt{s_{_{NN}}} = $200 GeV}

\newcommand{\abilene}{Abilene Christian University, Abilene, Texas 79699, USA}
\newcommand{\banaras}{Department of Physics, Banaras Hindu University, Varanasi 221005, India}
\newcommand{\barc}{Bhabha Atomic Research Centre, Bombay 400 085, India}
\newcommand{\bnlcoll}{Collider-Accelerator Department, Brookhaven National Laboratory, Upton, New York 11973-5000, USA}
\newcommand{\bnlphys}{Physics Department, Brookhaven National Laboratory, Upton, New York 11973-5000, USA}
\newcommand{\caucr}{University of California - Riverside, Riverside, California 92521, USA}
\newcommand{\charlesczech}{Charles University, Ovocn\'{y} trh 5, Praha 1, 116 36, Prague, Czech Republic}
\newcommand{\chonbuk}{Chonbuk National University, Jeonju, 561-756, Korea}
\newcommand{\ciae}{China Institute of Atomic Energy (CIAE), Beijing, People's Republic of China}
\newcommand{\cns}{Center for Nuclear Study, Graduate School of Science, University of Tokyo, 7-3-1 Hongo, Bunkyo, Tokyo 113-0033, Japan}
\newcommand{\colorado}{University of Colorado, Boulder, Colorado 80309, USA}
\newcommand{\columbia}{Columbia University, New York, New York 10027 and Nevis Laboratories, Irvington, New York 10533, USA}
\newcommand{\czechtech}{Czech Technical University, Zikova 4, 166 36 Prague 6, Czech Republic}
\newcommand{\dapnia}{Dapnia, CEA Saclay, F-91191, Gif-sur-Yvette, France}
\newcommand{\debrecen}{Debrecen University, H-4010 Debrecen, Egyetem t{\'e}r 1, Hungary}
\newcommand{\elte}{ELTE, E{\"o}tv{\"o}s Lor{\'a}nd University, H - 1117 Budapest, P{\'a}zm{\'a}ny P. s. 1/A, Hungary}
\newcommand{\ewha}{Ewha Womans University, Seoul 120-750, Korea}
\newcommand{\fit}{Florida Institute of Technology, Melbourne, Florida 32901, USA}
\newcommand{\fsu}{Florida State University, Tallahassee, Florida 32306, USA}
\newcommand{\gsu}{Georgia State University, Atlanta, Georgia 30303, USA}
\newcommand{\hiroshima}{Hiroshima University, Kagamiyama, Higashi-Hiroshima 739-8526, Japan}
\newcommand{\ihepprot}{IHEP Protvino, State Research Center of Russian Federation, Institute for High Energy Physics, Protvino, 142281, Russia}
\newcommand{\illuiuc}{University of Illinois at Urbana-Champaign, Urbana, Illinois 61801, USA}
\newcommand{\inrras}{Institute for Nuclear Research of the Russian Academy of Sciences, prospekt 60-letiya Oktyabrya 7a, Moscow 117312, Russia}
\newcommand{\instpasczech}{Institute of Physics, Academy of Sciences of the Czech Republic, Na Slovance 2, 182 21 Prague 8, Czech Republic}
\newcommand{\isu}{Iowa State University, Ames, Iowa 50011, USA}
\newcommand{\jinrdubna}{Joint Institute for Nuclear Research, 141980 Dubna, Moscow Region, Russia}
\newcommand{\jyvaskyla}{Helsinki Institute of Physics and University of Jyv{\"a}skyl{\"a}, P.O.Box 35, FI-40014 Jyv{\"a}skyl{\"a}, Finland}
\newcommand{\kek}{KEK, High Energy Accelerator Research Organization, Tsukuba, Ibaraki 305-0801, Japan}
\newcommand{\kfki}{KFKI Research Institute for Particle and Nuclear Physics of the Hungarian Academy of Sciences (MTA KFKI RMKI), H-1525 Budapest 114, POBox 49, Budapest, Hungary}
\newcommand{\korea}{Korea University, Seoul, 136-701, Korea}
\newcommand{\kurchatov}{Russian Research Center ``Kurchatov Institute", Moscow, 123098 Russia}
\newcommand{\kyoto}{Kyoto University, Kyoto 606-8502, Japan}
\newcommand{\labllr}{Laboratoire Leprince-Ringuet, Ecole Polytechnique, CNRS-IN2P3, Route de Saclay, F-91128, Palaiseau, France}
\newcommand{\lawllnl}{Lawrence Livermore National Laboratory, Livermore, California 94550, USA}
\newcommand{\losalamos}{Los Alamos National Laboratory, Los Alamos, New Mexico 87545, USA}
\newcommand{\lpc}{LPC, Universit{\'e} Blaise Pascal, CNRS-IN2P3, Clermont-Fd, 63177 Aubiere Cedex, France}
\newcommand{\lund}{Department of Physics, Lund University, Box 118, SE-221 00 Lund, Sweden}
\newcommand{\maryland}{University of Maryland, College Park, Maryland 20742, USA}
\newcommand{\mass}{Department of Physics, University of Massachusetts, Amherst, Massachusetts 01003-9337, USA }
\newcommand{\muenster}{Institut fur Kernphysik, University of Muenster, D-48149 Muenster, Germany}
\newcommand{\muhlenberg}{Muhlenberg College, Allentown, Pennsylvania 18104-5586, USA}
\newcommand{\myongji}{Myongji University, Yongin, Kyonggido 449-728, Korea}
\newcommand{\nagasaki}{Nagasaki Institute of Applied Science, Nagasaki-shi, Nagasaki 851-0193, Japan}
\newcommand{\newmex}{University of New Mexico, Albuquerque, New Mexico 87131, USA }
\newcommand{\nmsu}{New Mexico State University, Las Cruces, New Mexico 88003, USA}
\newcommand{\ornl}{Oak Ridge National Laboratory, Oak Ridge, Tennessee 37831, USA}
\newcommand{\orsay}{IPN-Orsay, Universite Paris Sud, CNRS-IN2P3, BP1, F-91406, Orsay, France}
\newcommand{\peking}{Peking University, Beijing, People's Republic of China}
\newcommand{\pnpi}{PNPI, Petersburg Nuclear Physics Institute, Gatchina, Leningrad region, 188300, Russia}
\newcommand{\riken}{RIKEN Nishina Center for Accelerator-Based Science, Wako, Saitama 351-0198, Japan}
\newcommand{\rikjrbrc}{RIKEN BNL Research Center, Brookhaven National Laboratory, Upton, New York 11973-5000, USA}
\newcommand{\rikkyo}{Physics Department, Rikkyo University, 3-34-1 Nishi-Ikebukuro, Toshima, Tokyo 171-8501, Japan}
\newcommand{\saispbstu}{Saint Petersburg State Polytechnic University, St. Petersburg, 195251 Russia}
\newcommand{\saopaulo}{Universidade de S{\~a}o Paulo, Instituto de F\'{\i}sica, Caixa Postal 66318, S{\~a}o Paulo CEP05315-970, Brazil}
\newcommand{\seoulnat}{Seoul National University, Seoul, Korea}
\newcommand{\stonybrkc}{Chemistry Department, Stony Brook University, SUNY, Stony Brook, New York 11794-3400, USA}
\newcommand{\stonycrkp}{Department of Physics and Astronomy, Stony Brook University, SUNY, Stony Brook, New York 11794-3400, USA}
\newcommand{\tenn}{University of Tennessee, Knoxville, Tennessee 37996, USA}
\newcommand{\titech}{Department of Physics, Tokyo Institute of Technology, Oh-okayama, Meguro, Tokyo 152-8551, Japan}
\newcommand{\tsukuba}{Institute of Physics, University of Tsukuba, Tsukuba, Ibaraki 305, Japan}
\newcommand{\vandy}{Vanderbilt University, Nashville, Tennessee 37235, USA}
\newcommand{\waseda}{Waseda University, Advanced Research Institute for Science and Engineering, 17 Kikui-cho, Shinjuku-ku, Tokyo 162-0044, Japan}
\newcommand{\weizmann}{Weizmann Institute, Rehovot 76100, Israel}
\newcommand{\yonsei}{Yonsei University, IPAP, Seoul 120-749, Korea}
\affiliation{\abilene}
\affiliation{\banaras}
\affiliation{\barc}
\affiliation{\bnlcoll}
\affiliation{\bnlphys}
\affiliation{\caucr}
\affiliation{\charlesczech}
\affiliation{\chonbuk}
\affiliation{\ciae}
\affiliation{\cns}
\affiliation{\colorado}
\affiliation{\columbia}
\affiliation{\czechtech}
\affiliation{\dapnia}
\affiliation{\debrecen}
\affiliation{\elte}
\affiliation{\ewha}
\affiliation{\fit}
\affiliation{\fsu}
\affiliation{\gsu}
\affiliation{\hiroshima}
\affiliation{\ihepprot}
\affiliation{\illuiuc}
\affiliation{\inrras}
\affiliation{\instpasczech}
\affiliation{\isu}
\affiliation{\jinrdubna}
\affiliation{\jyvaskyla}
\affiliation{\kek}
\affiliation{\kfki}
\affiliation{\korea}
\affiliation{\kurchatov}
\affiliation{\kyoto}
\affiliation{\labllr}
\affiliation{\lawllnl}
\affiliation{\losalamos}
\affiliation{\lpc}
\affiliation{\lund}
\affiliation{\maryland}
\affiliation{\mass}
\affiliation{\muenster}
\affiliation{\muhlenberg}
\affiliation{\myongji}
\affiliation{\nagasaki}
\affiliation{\newmex}
\affiliation{\nmsu}
\affiliation{\ornl}
\affiliation{\orsay}
\affiliation{\peking}
\affiliation{\pnpi}
\affiliation{\riken}
\affiliation{\rikjrbrc}
\affiliation{\rikkyo}
\affiliation{\saispbstu}
\affiliation{\saopaulo}
\affiliation{\seoulnat}
\affiliation{\stonybrkc}
\affiliation{\stonycrkp}
\affiliation{\tenn}
\affiliation{\titech}
\affiliation{\tsukuba}
\affiliation{\vandy}
\affiliation{\waseda}
\affiliation{\weizmann}
\affiliation{\yonsei}
\author{A.~Adare} \affiliation{\colorado}
\author{S.~Afanasiev} \affiliation{\jinrdubna}
\author{C.~Aidala} \affiliation{\mass}
\author{N.N.~Ajitanand} \affiliation{\stonybrkc}
\author{Y.~Akiba} \affiliation{\riken} \affiliation{\rikjrbrc}
\author{H.~Al-Bataineh} \affiliation{\nmsu}
\author{J.~Alexander} \affiliation{\stonybrkc}
\author{K.~Aoki} \affiliation{\kyoto} \affiliation{\riken}
\author{Y.~Aramaki} \affiliation{\cns}
\author{E.T.~Atomssa} \affiliation{\labllr}
% ...
\collaboration{PHENIX Collaboration} \noaffiliation

\date{\today}

\begin{abstract}

Heavy quarkonia are observed to be suppressed in relativistic heavy ion 
collisions relative to their production in $p+p$ collisions scaled by the 
number of binary collisions.  In order to determine if this suppression is 
related to color screening of these states in the produced medium, one 
needs to account for other nuclear modifications including those in cold 
nuclear matter.  In this paper, we present new measurements from the 
PHENIX 2007 data set of $J/\psi$ yields at forward rapidity ($1.2 < |y| < 
2.2$) in Au$+$Au collisions at $\sqrt{s_{_{NN}}} = 200$ GeV.  The data 
confirm the earlier finding that the suppression of $J/\psi$ at forward 
rapidity is stronger than at midrapidity, while also extending the 
measurement to finer bins in collision centrality and higher transverse 
momentum ($p_T$).  We compare the experimental data to the most recent 
theoretical calculations that incorporate a variety of physics mechanisms 
including gluon saturation, gluon shadowing, initial-state parton energy 
loss, cold nuclear matter breakup, color screening, and charm 
recombination.  We find $J/\psi$ suppression beyond cold-nuclear-matter 
effects.  However, the current level of disagreement between models and 
$d+$Au data precludes using these models to quantify the 
hot-nuclear-matter suppression.

\end{abstract}

\pacs{25.75.Dw}



\maketitle


Heavy quarkonia have long been proposed as a sensitive probe of the 
color screening length and deconfinement in the quark-gluon 
plasma~\cite{Matsui:1986}. The picture that was originally proposed is 
complicated by other competing effects that modify quarkonia production 
and survival in cold and hot nuclear matter.  The large suppression of 
$\jpsi$ in \pbpb\ collisions at $\sqrts$ = 17.2 GeV measured by the 
NA50 experiment indicated suppression beyond that projected from cold 
nuclear matter effects and led to the initial conclusion that color 
screening was the dominant 
mechanism~\cite{Abreu:2000p3069,Kharzeev:1996yx,Heinz:2000bk}.  The 
expectation was that at the Relativistic Heavy Ion Collider (RHIC), 
where higher energy densities and temperatures are created, the $\jpsi$ 
suppression would be stronger and turn on in even more peripheral 
collisions.  However, measurements from the PHENIX experiment's 2004 
data set in \auau\ collisions at $\sqrts$ = 200 GeV revealed that the 
centrality-dependent nuclear modification factor $\raa$ at midrapidity 
was the same within statistical and systematic uncertainties as the 
NA50 result~\cite{PPG068,Nagle:2007zt,PereiraDaCosta:2005xz}.  In 
addition, for more central collisions ($\npart\gtrsim100$) the 
suppression was measured to be larger at forward rapidity ($1.2 < |y| < 
2.2$) compared with midrapidity~\cite{PPG068}.  This is opposite to the 
expectation that the suppression should be less at forward rapidity, 
where the energy density is lower.

The initial estimates of cold nuclear matter (CNM) effects in the NA50 
\pbpb\ data were based on \pa\ measurements~\cite{Alessandro:2004p3275} 
at higher collision energies. More recently, measurements of $\jpsi$ 
production in \pa\ collisions at $\sqrts$ = 17.2 GeV made by NA60 have 
shown that CNM effects are stronger than the initial estimates, due to 
a substantially larger $\jpsi$ effective breakup cross 
section~\cite{Arnaldi:2009ph}. This resulted in a reduction of the 
estimated suppression due to possible hot nuclear matter effects from 
$\sim$50\% to $\sim$25\%, relative to cold nuclear matter effects. 
Based on a systematic study of the energy 
dependence~\cite{Lourenco:2008sk} the breakup cross section is expected 
to be much smaller at the higher RHIC energies, leading to smaller 
overall CNM effects. Thus it now seems that at RHIC the suppression 
beyond cold nuclear matter effects at midrapidity could be higher 
compared to the NA50 results, as one would expect due to the higher 
energy density. However, the question of why the observed suppression 
at RHIC is stronger at forward rapidity than at midrapidity remains 
less understood.

Since that first PHENIX measurement, many alternative explanations have 
been proposed which now require rigorous confrontation with the full 
set of experimental measurements.  In this paper, we detail the 
measurement of forward rapidity $\jpsi$ yields and modifications in 
\auau\ collisions at $\sqrts$ = 200 GeV from data taken by the PHENIX 
experiment in 2007.  The data set is more than three times larger than 
the earlier published 2004 data set.  In addition, significant 
improvements in our understanding of the detector performance and 
signal extraction have led to a reduction in systematic uncertainties. 
We present details of this new analysis as well as comparisons with 
theoretical calculations that have attempted to reconcile the earlier 
data in terms of competing physics mechanisms.

\section{Data Analysis}

The PHENIX experiment is described in detail in~\cite{Adcox:2003p2584}.  
For the forward rapidity $\jpsi$ data analysis presented here, the 
PHENIX experiment utilizes one global detector, the beam-beam counter 
(BBC), for event centrality characterization and $z$-vertex 
determination, and two muon spectrometers (North and South) for 
measuring the $\jpsi$ via the dimuon decay channel.  The BBC is 
described in detail in~\cite{Allen:2003p2581}.  It comprises two arrays 
of 64 quartz \v{C}erenkov counters that measure charged particles 
within the pseudorapidity range ($3.0 < |\eta| < 3.9$).  The BBC is 
also used as the primary Level-1 trigger for \auau\ minimum bias 
events.  The two muon spectrometers are described in detail 
in~\cite{Akikawa:2003p2580}, and comprise an initial hadronic absorber 
followed by three sets of cathode strip chambers in a magnetic field, 
referred to as the Muon Tracker (MuTR).  Finally, there are five planes 
of active Iarocci tubes interleaved with additional steel absorber 
plates, referred to as the Muon Identifier (MuID).  The muon 
spectrometers measure $\jpsi$s over the rapidity range 
$1.2 < |y| < 2.2$.

The PHENIX data acquisition system is capable of recording minimum bias 
\auau\ collisions at high rates ($> 5$ kHz) with a data archival rate 
in excess of 600 MB/s. During the 2007 \auau\ run at $\sqrts$ = 200 
GeV, the experiment recorded 82\% of all collisions where the minimum 
bias level-1 trigger fired. Therefore no additional muon-specific 
trigger was necessary.  After data quality cuts to remove runs where 
there were significant detector performance variations, we analyzed 
$3.6 \times 10^{9}$ minimum bias \auau\ events.  The BBC Level-2 
trigger used as the minimum bias trigger for \auau\ events required at 
least two hits in each of the BBC arrays and a fast-reconstructed 
$z$-vertex within $\pm30$ cm of the nominal center of the detector.  
An additional selection for our minimum bias definition in offline 
reconstruction includes a requirement of at least one neutron hit in 
each of our Zero Degree Calorimeters (ZDCs).  This removes a 1-2\% 
background event contamination in the peripheral event sample.  The 
minimum bias definition corresponds to 92\%~$\pm$~3\% of the inelastic 
\auau\ cross section~\cite{PPG068}.  We further categorize the events 
in terms of centrality classes by comparing the combined North and 
South BBC charge to a negative binomial distribution for the number of 
produced particles within the BBC acceptance combined with a Glauber 
model of the collision~\cite{Miller:2007p723}.  For each centrality 
category, the mean number of participating nucleons ($\npart$), binary 
collisions ($\ncoll$), and impact parameter ($b$), and their associated 
systematic uncertainties are shown in Table~\ref{table:npart_ncoll}.

\begin{table}[tb!]
\begin{center}
  \caption{Mean $\npart$, $\ncoll$, and impact parameter values 
and systematic uncertainties in each centrality category.}
  \label{table:npart_ncoll}
\begin{ruledtabular}
\begin{tabular}{cccc}
Centrality (\%) & $\langle\npart\rangle$ & $\langle\ncoll\rangle$ &  
$\langle b\rangle$ (fm)\\
\hline
0--5   & 350.8$\pm$3.1 & 1067.0$\pm$107.7  & 2.3$\pm$0.1 \\
5--10  & 301.7$\pm$4.7 & 857.8$\pm$85.5 & 3.9$\pm$0.1 \\
10--15 & 255.7$\pm$5.4 & 680.2$\pm$67.3 & 5.2$\pm$0.2 \\
15--20 & 216.4$\pm$5.6 & 538.7$\pm$52.4 & 6.1$\pm$0.2  \\
20--25 & 182.4$\pm$5.7 & 424.4$\pm$40.4 & 7.0$\pm$0.3  \\
25--30 & 152.7$\pm$5.9 & 330.9$\pm$32.7 & 7.7$\pm$0.3 \\
30--35 & 126.8$\pm$5.9 & 254.7$\pm$25.8 & 8.4$\pm$0.3 \\
35--40 & 104.2$\pm$5.8 & 193.1$\pm$20.7 & 9.0$\pm$0.3 \\
40--45 & 84.6$\pm$5.6 & 143.9$\pm$16.5 & 9.6$\pm$0.4 \\
45--50 & 67.7$\pm$5.4 & 105.4$\pm$13.5 & 10.2$\pm$0.4 \\
50--55 & 53.2$\pm$5.0 & 75.2$\pm$10.5 & 10.7$\pm$0.4 \\
55--60 & 41.0$\pm$4.5 & 52.5$\pm$8.2 & 11.2$\pm$0.4 \\
60--65 & 30.8$\pm$3.9 & 35.7$\pm$6.1 & 11.7$\pm$0.5 \\
65--70 & 22.6$\pm$3.4 & 23.8$\pm$4.7 & 12.2$\pm$0.5 \\
70--75 & 16.1$\pm$2.8 & 15.4$\pm$3.3 & 12.6$\pm$0.5 \\
75--80 & 11.2$\pm$2.2 & 9.7$\pm$2.3 & 13.1$\pm$0.5 \\
80--92 & 5.6$\pm$0.8 & 4.2$\pm$0.8 & 13.9$\pm$0.5 \\
\\
0--20  & 280.5$\pm$4.6 & 783.2$\pm$77.5  & 4.4$\pm$0.2 \\
20--40 & 141.5$\pm$5.8 & 300.8$\pm$29.6 & 8.0$\pm$0.3 \\
40--60 & 61.6$\pm$5.1 & 94.2$\pm$12.0   & 10.4$\pm$0.4 \\
60--92 & 14.4$\pm$2.1 & 14.5$\pm$2.8   & 13.0$\pm$0.5 \\
\end{tabular} \end{ruledtabular}
\end{center}
\end{table}

From this minimum bias data sample, we reconstruct muon candidates by 
finding tracks that penetrate through all layers of the MuID, then 
matching to tracks in the MuTR.  The requirement of the track 
penetrating the full absorber material through the MuID significantly 
reduces the hadron contribution. However, there is a small probability 
of order $\sim$1/1000 for a charged hadron to penetrate the material 
without suffering a hadronic interaction. This is referred to as a 
punch-through hadron.  Additionally, the current muon spectrometer 
cannot reject most muons that originate from charged pions and kaons 
which decay before the absorber in front of the MuTR.  Pairs of muon 
candidate tracks are selected and a combined fit is performed with the 
collision $z$-vertex from the BBC.  We apply various cuts to enhance 
the sample of good muon track pairs, including cuts on the individual 
track $\chi^2$ values, the matching between position and direction 
vectors of the MuID track and the MuTR track projected to the front of 
the MuID, and finally the track pair and BBC $z$-vertex combined fit 
$\chi^2_{\rm vtx}$.

We then calculate the invariant mass of all muon candidate unlike 
charged sign pairs in various bins in rapidity, $\pt$, and centrality. 
Due to the high particle multiplicity in \auau\ events, there is a 
significant background under the $\jpsi$ peak.  In the 0-5\% most 
central \auau\ collisions, within the mass window around the $\jpsi$ 
($2.6 < M$ [GeV/$c^2$] $ < 3.6$) the signal to background is of order 
2.9\% in the South muon spectrometer and 0.7\% in the North muon 
spectrometer (which has a different geometric acceptance and 
significantly higher occupancy).  The background comprises two 
components.  First, there is the combinatorial background from 
uncorrelated track pairs.  Second, there is a correlated background 
from physical sources including open charm pair decays (e.g.~$D^{0} + 
\overline{D^0} \rightarrow (K^- \mu^+ \nu_{\mu})
+ (K^+ \mu^- \overline{{\nu}_{\mu}}) $, 
open beauty pair decays, and Drell-Yan.  The combinatorial background is 
estimated and subtracted by using event mixing to recreate the 
background from uncorrelated pairs.  Pairs are created from different 
\auau\ events within the same category in \auau\ centrality and BBC 
collision z-vertex.  The mixed-event invariant mass distributions are 
calculated for unlike-sign and like-sign pairs.  We treat the real 
event like-sign pairs as being purely from combinatorial background, 
since the contribution of the above-mentioned correlated physical 
background is negligible relative to the combinatorial background.  
Thus, we determine the mixed event normalization for the unlike-sign 
case by calculating the normalization factor between the mixed event 
and real event like-sign counts.  We have confirmed that the mixed 
event and real event like-sign invariant mass distributions match over 
the full mass range used in the analysis.

\begin{figure*}[tbh]
\centering
\includegraphics[width=0.48\linewidth]{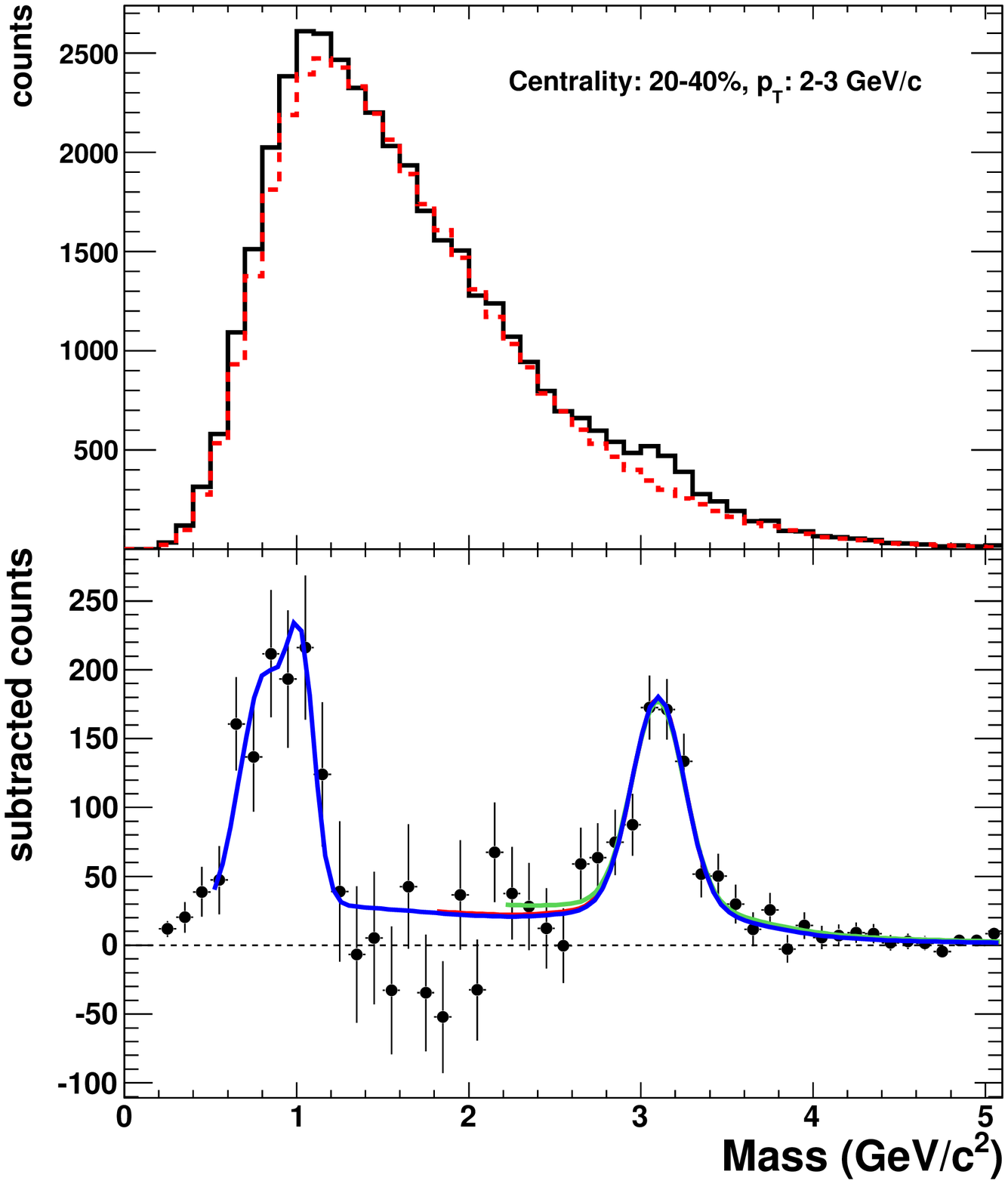}
\includegraphics[width=0.48\linewidth]{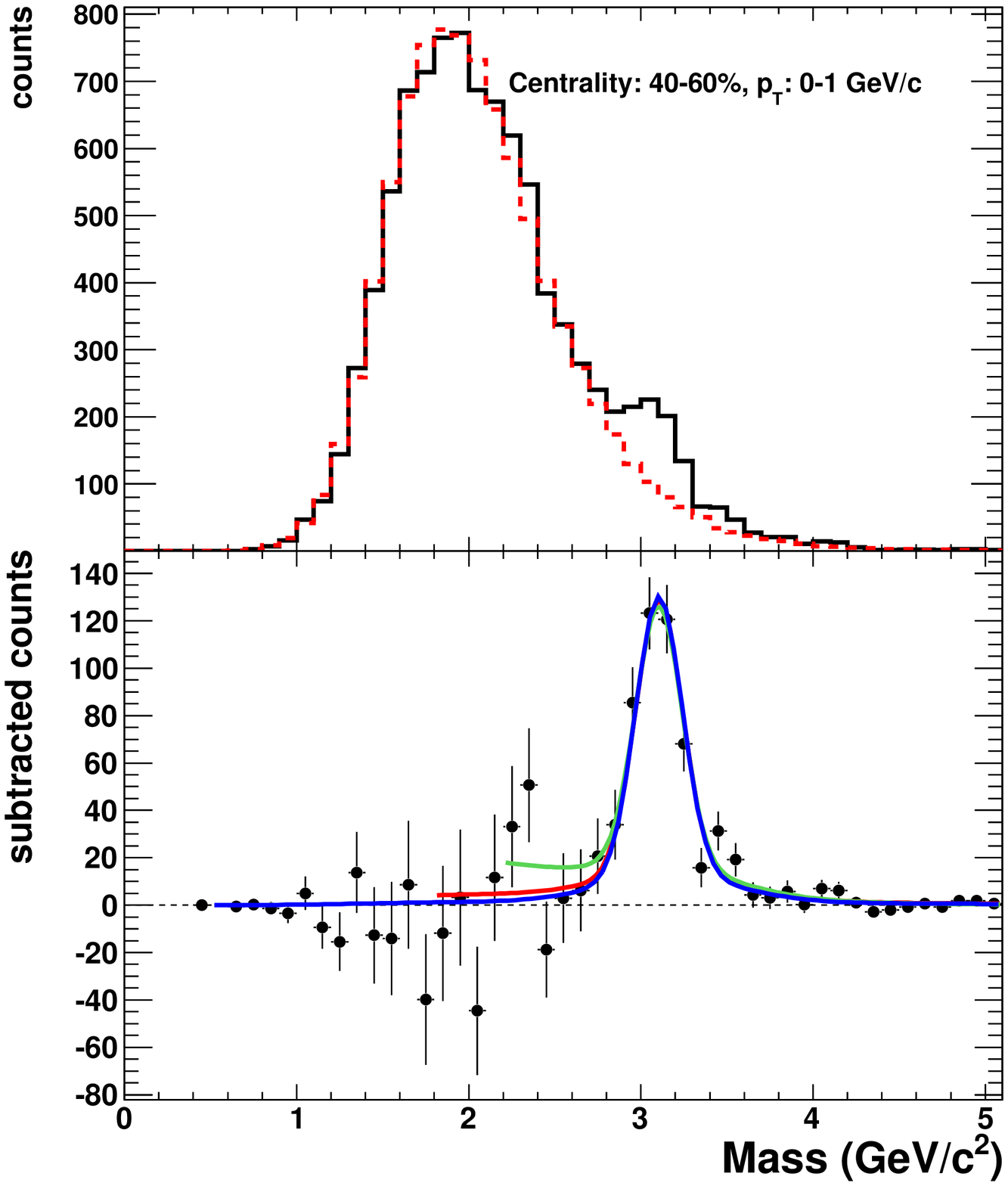}
\caption{(Color online) Top: muon spectrometer unlike-sign invariant
mass distributions from same-event pairs (black, solid) and mixed-event
pairs (red, dashed).  Two example centralities and transverse
momentum ranges are shown.  Bottom: Corresponding unlike-sign
invariant mass distributions after mixed-event 
combinatorial-background subtraction.  Colored curves represent 
three different mass ranges for the signal extraction fit.}
\label{fig:mass_plots}
\end{figure*}

The total $\jpsi$ counts recorded in all \auau\ collisions are 
$\sim$9100 and $\sim$4900 in the south and north muon spectrometers, 
respectively.  Two example unlike-sign invariant mass distributions 
before (upper panel) and after (lower panel) mixed-event combinatorial 
background subtraction are shown in Fig.~\ref{fig:mass_plots}.  The 
number of measured $\jpsi$ is derived from the subtracted spectra by 
fitting to the data the line shape of the $\jpsi$, as determined in 
$\pp$ collisions~\cite{Adare:2007gn}, and an exponential for the 
remaining correlated physical background.  Note that in the higher $\pt 
> 2$ GeV/$c$ bins, there is some small acceptance for the $\rho, 
\omega$ and $\phi$ and thus these additional components are included in 
the fit.  At low $\pt$, the acceptance goes to zero for lower invariant 
mass due to the minimum required momentum for each muon to penetrate the 
MuID and the angular acceptance of the spectrometer.  This is accounted 
for by folding the fit function with an acceptance function that is 
calculated from Monte Carlo simulation and goes to zero at small invariant 
mass, as expected.  We perform a set of fits where we vary the invariant 
mass range, the line shape of the $\jpsi$, and the normalization of the 
mixed-event sample by $\pm 2\%$ to determine the mean extracted $\jpsi$ 
signal and systematic uncertainty from the RMS of the different results. 
Note that any bin where the extracted signal is of less than one standard 
deviation significance (including statistical and systematic 
uncertainties) is quoted as a 90\% confidence level upper limit (CLUL) 
based on Poisson statistics.

We also estimated the combinatorial background using the like-sign 
method, i.e.~$N_{+-}^{comb}=2\sqrt{N_{++} \times N_{--}}$.  In this 
case there is no event mixing, the background is estimated purely from 
same-event like-sign pairs (instead of mixed-event pairs).  The two 
methods agree over most of the centrality range; however, for the more 
central events the like-sign method results in somewhat lower extracted 
counts ($\sim 10\%$).  We take the average of the two signal extraction 
methods and assign an additional systematic uncertainty due to the 
difference, although for peripheral bins it is a negligible difference.

We then calculate the $\jpsi$ invariant yield for each centrality bin
and also in bins in $\pt$ by the following equation:
\begin{equation}
B_{\mu\mu} \frac{d^3N}{d\pt^2 dy} = 
\frac{1}{2\pi\pt{\scriptstyle\Delta}\pt{\scriptstyle\Delta} y}
\frac{N_{\jpsi}}{A\epsilon\ N_{EVT}}
\end{equation}
where $B_{\mu\mu}$ is the branching fraction of $\jpsi$ to muons, 
$N_{\jpsi}$ is the number of measured $\jpsi$, $N_{EVT}$ is the number 
of events in the relevant \auau\ centrality category, $A\epsilon$ is 
the detector geometric acceptance and efficiency, and 
${\scriptstyle\Delta}\pt$ and ${\scriptstyle\Delta}y$ are the bin width 
in $\pt$ and $y$.  For the $\pt$-integrated bins, we similarly 
calculate $B_{\mu\mu}dN/dy = N_{\jpsi}/( A\epsilon N_{EVT} 
{\scriptstyle\Delta} y)$.

\begin{figure}[tbh]
\centering
\includegraphics[width=1.0\linewidth]{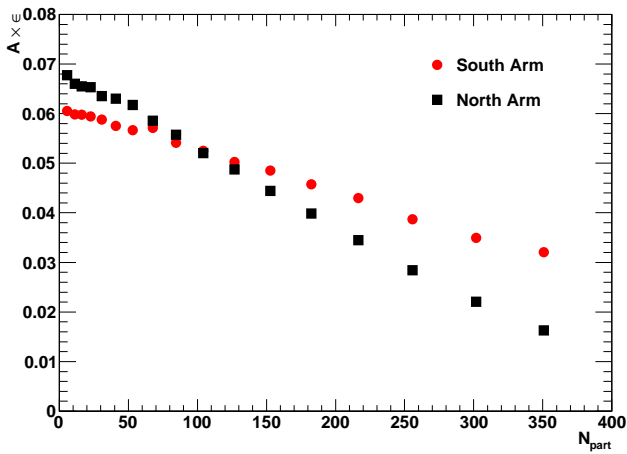}
\caption{(Color online) Acceptance $\times$ efficiency as a function
of centrality for the North and South Arms.}\label{fig:acceff}
\end{figure}

We calculate $A\epsilon$ to correct for the geometric acceptance of the 
detector and the inefficiencies of the MuTR and MuID, the track finding 
algorithm, and occupancy-related effects in the \auau\ environment.  
This is done by propagating {\sc pythia}-generated $\jpsi$ through the 
PHENIX {\sc geant}-3~\cite{geant3} detector simulation, and embedding 
the resulting hits in real \auau\ events.  The events are then 
reconstructed using the identical analysis as for real data, and the 
ratio is taken between reconstructed and embedded $\jpsi$.  The 
resulting $A\epsilon$ as a function of centrality is shown in 
Fig.~\ref{fig:acceff} for both the North and South Muon Arms.  The 
effect of the detector occupancy can be seen for more central events, 
as well as the higher occupancy and resulting lower efficiency in the 
North Arm.

The acceptance and efficiency as a function of $\pt$ is relatively 
flat, with a modest 20\% decrease from $\pt$ = 0 to $\pt\sim$2.5 
GeV/$c^2$, then proceeds to rise again.  The behavior is essentially 
the same across centralities, with the only difference being the 
absolute scale of $A\epsilon$.

We calculate the invariant yields separately from the two muon 
spectrometers, and then combine the values for the final results.  We 
take the weighted average based on the statistical uncertainties and 
those systematic uncertainties which are uncorrelated between the two 
measurements.  The final averaged result is assigned the uncorrelated 
(reduced by the averaging) and correlated systematic uncertainties. In 
addition, it was found that the invariant yields from the two 
spectrometers disagree beyond their independent uncertainties, and a 
5\% systematic was added to account for the difference.

\section{Results}

The final $\jpsi$ invariant yields in 5\%-wide centrality bins 
(integrated over all $\pt$) are listed in 
Table~\ref{table:invyield_cent_avg}, and as a function of $\pt$ in 
broader 20\%-wide centrality bins in Table~\ref{table:invyield_pt_avg}.

\begin{figure}[htb]
\centering
\includegraphics[width=1.0\linewidth]{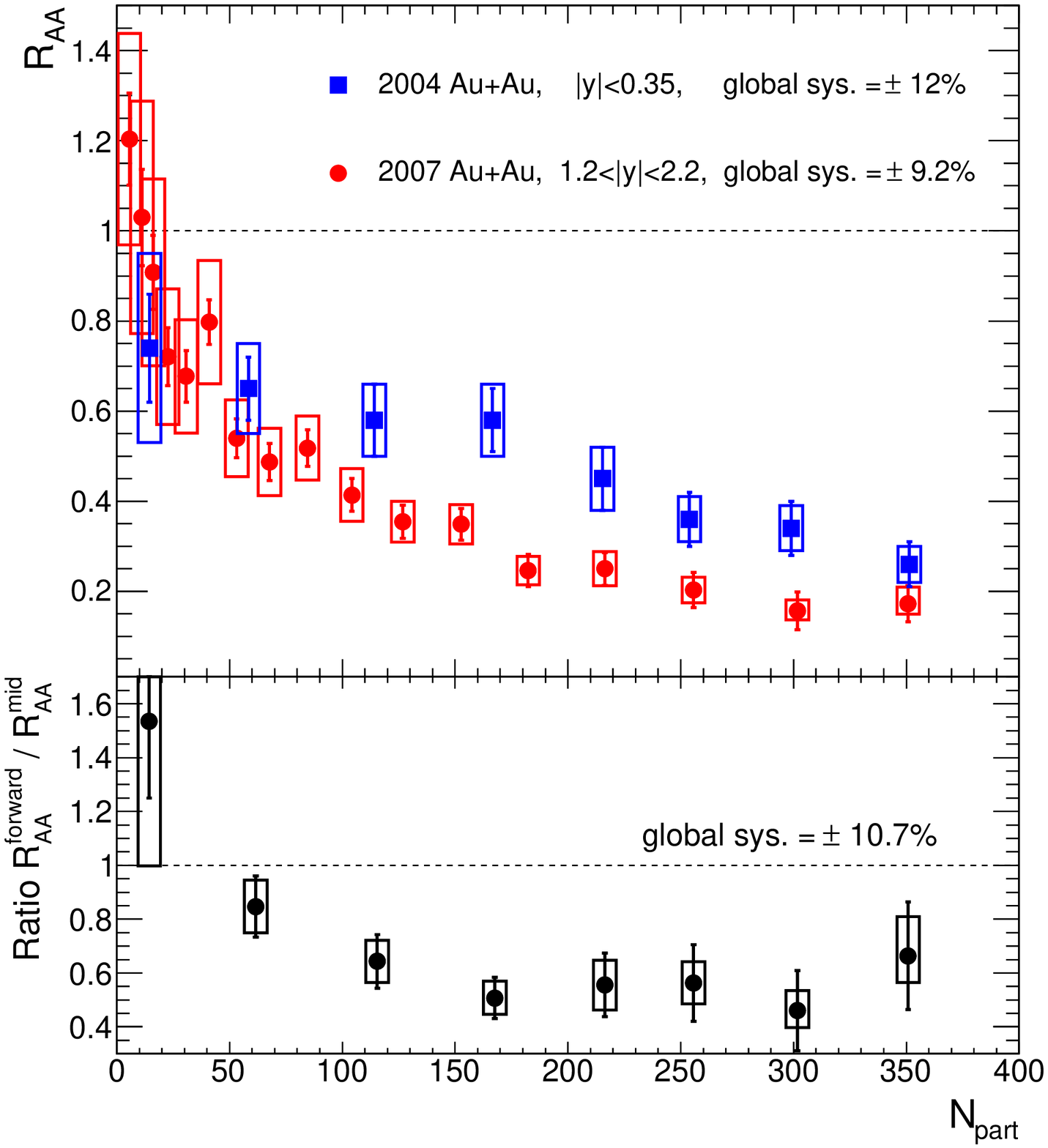}
\caption{(Color online) $\jpsi$ $\raa$ as a function of $\npart$.  
Error bars represent the statistical and uncorrelated systematic 
uncertainties, while the boxes represent the point-to-point correlated 
systematics. The global scale systematic uncertainties are quoted as 
text.  The lower panel contains the ratio of forward rapidity to 
midrapidity for the points in the upper panel.}\label{fig:raa_npart}
\end{figure}

\begingroup \squeezetable
\begin{table}[hb!]
\begin{center}
\caption{$\jpsi$ invariant yields $B_{\mu\mu} dN/dy$ at forward rapidity ($1.2<|y|<2.2$) vs. \auau\ collision centrality.  The type C (global) uncertainty for all points is 10.7\%.}
\label{table:invyield_cent_avg}
\begin{ruledtabular}
\begin{tabular*}{\linewidth}{@{\extracolsep{\fill}}cccccccccc}
Centrality & $B_{\mu\mu}dN/dy$ & $\pm$stat & $\pm$type A & $+$type B & 
$-$type B & scale \\
(\%) & \multicolumn{6}{c}{(Gev/$c^{-2}$)}\\
\hline
0--5 & 1.25 & 0.27 & 0.11 & 0.24 & 0.12 & $\times10^{-4}$ \\
5--10 & 9.12 & 2.46 & 0.13 & 1.15 & 0.81 & $\times10^{-5}$ \\
10--15 & 9.37 & 1.66 & 0.71 & 1.01 & 0.95 & $\times10^{-5}$ \\
15--20 & 9.16 & 1.24 & 0.48 & 1.08 & 1.08 & $\times10^{-5}$ \\
20--25 & 7.11 & 0.98 & 0.29 & 0.64 & 0.64 & $\times10^{-5}$ \\
25--30 & 7.85 & 0.74 & 0.29 & 0.64 & 0.64 & $\times10^{-5}$ \\
30--35 & 6.14 & 0.57 & 0.28 & 0.49 & 0.49 & $\times10^{-5}$ \\
35--40 & 5.43 & 0.45 & 0.14 & 0.52 & 0.52 & $\times10^{-5}$ \\
40--45 & 5.07 & 0.37 & 0.15 & 0.39 & 0.39 & $\times10^{-5}$ \\
45--50 & 3.49 & 0.28 & 0.07 & 0.30 & 0.30 & $\times10^{-5}$ \\
50--55 & 2.76 & 0.21 & 0.07 & 0.21 & 0.21 & $\times10^{-5}$ \\
55--60 & 2.85 & 0.17 & 0.04 & 0.21 & 0.21 & $\times10^{-5}$ \\
60--65 & 1.64 & 0.14 & 0.03 & 0.12 & 0.12 & $\times10^{-5}$ \\
65--70 & 1.17 & 0.10 & 0.01 & 0.09 & 0.09 & $\times10^{-5}$ \\
70--75 & 9.49 & 0.85 & 0.11 & 0.73 & 0.73 & $\times10^{-6}$ \\
75--80 & 6.79 & 0.69 & 0.14 & 0.51 & 0.51 & $\times10^{-6}$ \\
80--92 & 3.43 & 0.29 & 0.04 & 0.26 & 0.26 & $\times10^{-6}$ \\
\end{tabular*}
\end{ruledtabular}
\end{center}
\end{table}
\endgroup

\begingroup \squeezetable
\begin{table}[h]
\begin{center}
\caption{$\jpsi$ invariant yields $B_{\mu\mu}\frac{d^3N}{d\pt^2dy}$
at forward rapidity ($1.2<|y|<2.2$) vs.\@ $\pt$ in four 
bins of \auau\ collision centrality.  The type C (global) uncertainties 
are 10\%, 10\%, 13\%, and 19\% for 0-20\%, 20-40\%, 40-60\%, and 60-92\% 
centrality, respectively.  Bins in which the $\jpsi$ yield was less than 
the combined statistical and systematic uncertainties are calculated as 
90\% Confidence Level Upper Limits (CLUL).}
\label{table:invyield_pt_avg}
\begin{ruledtabular}
\begin{tabular*}{\linewidth}{@{\extracolsep{\fill}}cccccccc}
Centrality & $\pt$ & $B_{\mu\mu}d^3N$ & stat & type A & 
$+$B & $-$B & scale\\
(\%) & (GeV/c) & $\overline{d\pt^2dy}$ &
\multicolumn{5}{c}{(Gev/$c^{-2}$)} \\
\hline
0--20 & 0--1 & 9.36 & 1.41 & 0.74 & 1.01 & 1.01 & $\times10^{-6}$ \\
 & 1--2 & 4.46 & 0.66 & 0.23 & 0.40 & 0.40 & $\times10^{-6}$ \\
 & 2--3 & 1.37 & 0.29 & 0.08 & 0.17 & 0.17 & $\times10^{-6}$ \\
 & 3--4 & 2.99 & 1.12 & 0.09 & 0.27 & 0.27 & $\times10^{-7}$ \\
 & 4--5 & 2.05 & 0.43 & 0.14 & 0.18 & 0.18 & $\times10^{-7}$ \\
 & 5--6 & \multicolumn{6}{c}{90\% CLUL = 3.27 $\times10^{-8}$} \\
 & 6--7 & \multicolumn{6}{c}{90\% CLUL = 2.00 $\times10^{-8}$} \\
\\
20--40 & 0--1 & 5.08 & 0.54 & 0.18 & 0.67 & 0.67 & $\times10^{-6}$ \\
 & 1--2 & 2.78 & 0.22 & 0.09 & 0.26 & 0.26 & $\times10^{-6}$ \\
 & 2--3 & 1.11 & 0.10 & 0.02 & 0.09 & 0.09 & $\times10^{-6}$ \\
 & 3--4 & 2.76 & 0.34 & 0.11 & 0.25 & 0.25 & $\times10^{-7}$ \\
 & 4--5 & 7.47 & 1.37 & 0.31 & 1.35 & 1.35 & $\times10^{-8}$ \\
 & 5--6 & 2.68 & 0.61 & 0.08 & 0.31 & 0.31 & $\times10^{-8}$ \\
 & 6--7 & \multicolumn{6}{c}{90\% CLUL = 7.15 $\times10^{-9}$} \\
\\
40--60 & 0--1 & 3.19 & 0.21 & 0.06 & 0.26 & 0.26 & $\times10^{-6}$ \\
 & 1--2 & 1.49 & 0.09 & 0.03 & 0.12 & 0.12 & $\times10^{-6}$ \\
 & 2--3 & 4.80 & 0.38 & 0.11 & 0.39 & 0.39 & $\times10^{-7}$ \\
 & 3--4 & 1.27 & 0.13 & 0.02 & 0.11 & 0.11 & $\times10^{-7}$ \\
 & 4--5 & 3.86 & 0.49 & 0.02 & 0.41 & 0.41 & $\times10^{-8}$ \\
 & 5--6 & 7.51 & 1.69 & 0.05 & 2.04 & 2.04 & $\times10^{-9}$ \\
 & 6--7 & \multicolumn{6}{c}{90\% CLUL = 2.82 $\times10^{-9}$} \\
\\
60--92 & 0--1 & 9.05 & 0.57 & 0.07 & 0.73 & 0.73 & $\times10^{-7}$ \\
 & 1--2 & 3.40 & 0.22 & 0.04 & 0.27 & 0.27 & $\times10^{-7}$ \\
 & 2--3 & 9.19 & 0.91 & 0.16 & 0.75 & 0.75 & $\times10^{-8}$ \\
 & 3--4 & 2.21 & 0.35 & 0.04 & 0.20 & 0.20 & $\times10^{-8}$ \\
 & 4--5 & 8.13 & 1.39 & 0.01 & 0.70 & 0.70 & $\times10^{-9}$ \\
 & 5--6 & 2.31 & 0.54 & 0.00 & 0.40 & 0.40 & $\times10^{-9}$ \\
 & 6--7 & \multicolumn{6}{c}{90\% CLUL = 7.69 $\times10^{-10}$} \\
\end{tabular*} \end{ruledtabular}
\end{center}
\end{table}
\endgroup

The nuclear modification factor $\raa$ compares $\jpsi$ production in 
$\myaa$ with binary collision-scaled $\pp$ reactions, and is calculated 
as: \begin{equation} \raa = \frac{1}{\langle\ncoll\rangle} 
\frac{dN^{\myaa}/dy}{dN^{\pp}/dy} \end{equation} where 
$\langle\ncoll\rangle$ is the mean number of binary collisions in the 
centrality category of interest.  The $\pp$ results are from the 
combined analysis of data taken in 2006 and 2008 as published 
in~\cite{Adare:2010fn}.  The resulting $\raa$ as a function of $\npart$ 
for $\jpsi$ from \auau\ collisions is shown as red circles in 
Fig.~\ref{fig:raa_npart}.

\begin{figure}
\centering
\includegraphics[width=1.0\linewidth]{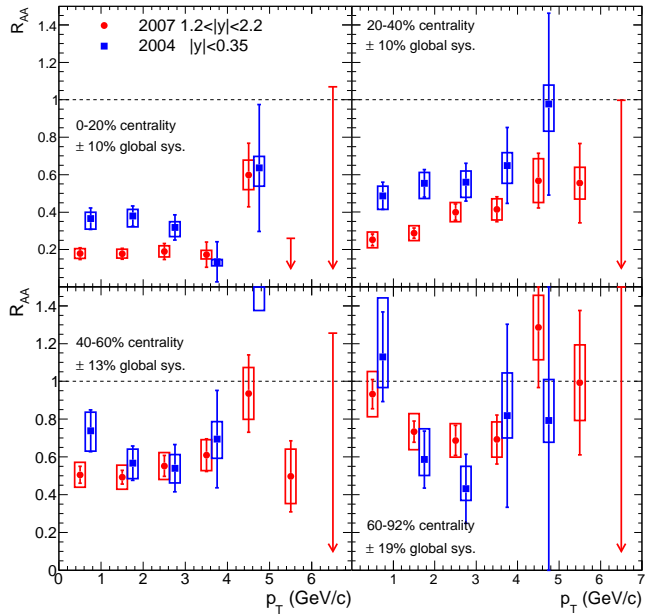}
\caption{(Color online) $\jpsi$ $\raa$ as a function of $\pt$ 
in four centrality bins.  Error bars represent the statistical and 
uncorrelated systematic uncertainties, while the boxes represent 
the point-to-point correlated systematic uncertainties.  The global
scale systematic uncertainties are quoted as
text.}\label{fig:raa_pt}
\end{figure}

The systematic uncertainties are divided into three categories: type A are 
the point-to-point uncorrelated systematics, type B are correlated (or 
anti-correlated) point-to-point, and type C are 100\% correlated (i.e.~a 
common multiplicative factor) between all of the points.  The error bars 
in Figs.~\ref{fig:raa_npart}--\ref{fig:raa_pt_zhao2} represent the 
statistical and type A uncertainties added in quadrature, the boxes 
represent the type B uncertainties, while the type C systematics are 
included as text in the labels.  The type A uncertainties are the RMS of 
the various mass fits as described above.  The type B uncertainties on 
$\raa$ are comprised of many sources, dominated by uncertainties in 
$\langle \ncoll \rangle$, uncertainties in the matching of Monte Carlo and 
real detector performance, and differences in signal extraction methods.  
The type C uncertainties are dominated by the normalization in the $\pp$ 
invariant cross section measurement. Important systematics on the 
invariant yields are listed in Table~\ref{table:systematics}.

\begin{table}[tb!]
\begin{center}
\caption{Systematic uncertainties on $dN/dy$ for central and peripheral centrality categories.}
\label{table:systematics}
\begin{ruledtabular}
\begin{tabular*}{\linewidth}{@{\extracolsep{\fill}}lccc}
Source & Central & Peripheral & Type \\
\hline
Signal extraction & 9.8\% & 1.3\% & A \\
Acceptance & 3.4\% & 2.2\% & B \\
Input $y$, $\pt$ distributions & 4\% & 4\% & B \\
Difference between mixed event/ & & & \\
like-sign background estimates & 1.5\% & 0.6\% & B \\
North/south arm agreement & 5\% & 5\% & B \\
MuID efficiency & 3.6\% & 2.8\% & B \\
\end{tabular*}
\end{ruledtabular}
\end{center}
\end{table}

For comparison, in Fig.~\ref{fig:raa_npart} we show our 
previously-published midrapidity $\jpsi$ $\raa$ values from data taken 
in 2004~\cite{PPG068}.  The midrapidity measurement was made in the 
PHENIX central spectrometers via the $\jpsi$ dielectron decay. There is 
no PHENIX updated measurement at midrapidity from the 2007 data set due 
to significantly increased conversion backgrounds from this engineering 
run of the PHENIX Hadron Blind Detector~\cite{Milov:2007xh}.  The ratio 
of the new forward rapidity data to the previously-published 
midrapidity data, shown in the lower panel of 
Fig.~\ref{fig:raa_npart}, is in agreement with the previous 
result~\cite{PPG068}, where the latter led to speculation as to what 
mechanism could cause a narrower rapidity distribution in \auau\ than 
$\pp$ collisions.

We also calculate $\raa$ as a function of $\pt$, again using the 
published 2006 and 2008 $\pp$ data~\cite{Adare:2010fn}.  Shown in 
Fig.~\ref{fig:raa_pt} are the new results at forward rapidity along 
with the previously-published 2004 midrapidity results~\cite{PPG068}. 
In some centrality bins for $\pt > 5$ GeV/$c$, we have no significant 
$\jpsi$ signal in \auau\ and thus can only quote a 90\% confidence 
level upper limit on $\raa$.

As there has been much recent interest in whether the $\raa$ as a 
function of $\pt$ rises or falls, we have performed a simple linear fit 
to the $\raa$ at forward rapidity over the full $\pt$ range and obtain 
the following slope ($m$) values:
$m=+0.011 \pm 0.018~c$/GeV (0-20\% central), 
$m=+0.065 \pm 0.023~c$/GeV (20-40\% central), 
$m=+0.034 \pm 0.033~c$/GeV (40-60\% central), and 
$m=-0.037 \pm 0.053~c$/GeV (60-92\% central).  
The quoted slope uncertainties are the quadrature sum of the 
statistical and systematic uncertainties.  Thus, only the 20-40\% 
centrality indicates a statistically significant increase in $\raa$ 
with $\pt$.

\section{Model Comparisons}

As previously mentioned, various theoretical models have been proposed 
to reconcile the $\jpsi$ suppression pattern previously 
published~\cite{PPG068}.  Here we compare our measurements with three 
calculations for the centrality and rapidity dependence of the 
suppression. The first deals entirely with initial-state effects, while 
the other two incorporate strong final-state effects.  Then we compare 
our measurements with a simple cold nuclear matter effect calculation 
extrapolated to \auau\ collisions.  Finally, we compare to a model 
calculation for the $\pt$ dependence of $\raa$ at forward rapidity.

In addition to the models discussed here, there are many more models 
that only have a midrapidity prediction for $\jpsi$ production.  
Because this paper is focused on forward rapidity $\jpsi$ production, 
we have not included comparisons to those models.

\subsection{Gluon Saturation Model}

In the first model by Kharzeev~et. al~\cite{Kharzeev:2009p2548}, it is 
assumed that the nuclear wave functions in very high-energy nuclear 
collisions can be described by the Color Glass Condensate (CGC).  The 
primary effect is the suppression of $\jpsi$ production and narrowing 
of the rapidity distribution due to saturation of the gluon fields in 
heavy ion collisions relative to $\pp$ collisions.  In addition, the 
production mechanism is modified from $\pp$ such that the multigluon 
exchange diagrams are enhanced.  It should be noted that this model 
does not include any hot medium effects, but does have a free parameter 
for the overall normalization factor for the \auau\ production, which 
is fixed to match the midrapidity central collision $\jpsi$ 
suppression.  Thus, the suppression trend with centrality and the 
relative suppression between mid and forward rapidity are predicted, 
but not the overall level of suppression.

\begin{figure}
\centering
\includegraphics[width=1.0\linewidth]{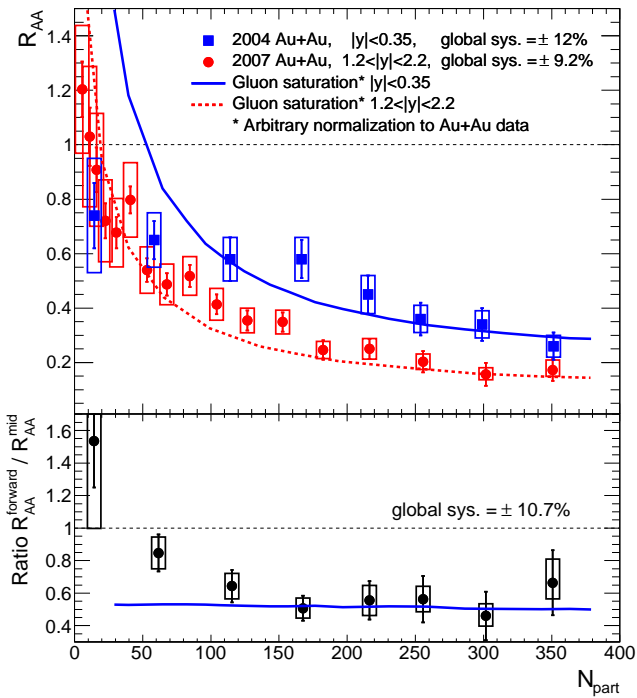}
\caption{(Color online) $\jpsi$ $\raa$ as a function of $\npart$.  
Model calculation from~\cite{Kharzeev:2009p2548}.  Lower panel is the 
ratio of forward to midrapidity points and curves from the upper
panel.}\label{fig:raa_npart_cgc}
\end{figure}

The resulting $\raa$ values calculated using this model are shown in 
Fig.~\ref{fig:raa_npart_cgc}.  This model provides a reasonable 
description of the data and in particular matches the observed larger 
suppression at forward rapidity than mid rapdity in central events 
($\raa^{\rm forward}/\raa^{\rm mid} \sim 0.5$).  It is notable that 
this ratio is essentially independent of centrality in their 
calculation, whereas the experimental data shows the relative 
suppression approaching one in the most peripheral events.  
Additionally, the calculation at midrapidity actually indicates a 
significant enhancement (i.e.~$\raa$ $> 1$) for peripheral events with 
$\npart < 50$.  
This enhancement is related to a coherence effect of double gluon 
exchange.  However, the coherence predicts an enhancement in \dau\ 
collisions and $\rdau$ at midrapidity, and no such enchancement is seen in 
the experimental data~\cite{Adare:2010fn}.

Recently, the appropriate normalization factor for the above CGC 
calculation has been calculated~\cite{Tuchin:2011,Nardi:2011}. 
Replacing the normalization factor previously applied to match the theory 
to the magnitude of the observed midrapidity suppression, results in a 
predicted CGC suppression approximately a factor of two smaller than in 
the /auau/ data.  This result suggests the importance of additional hot 
nuclear matter effects.

\begin{figure}
\centering
\includegraphics[width=1.0\linewidth]{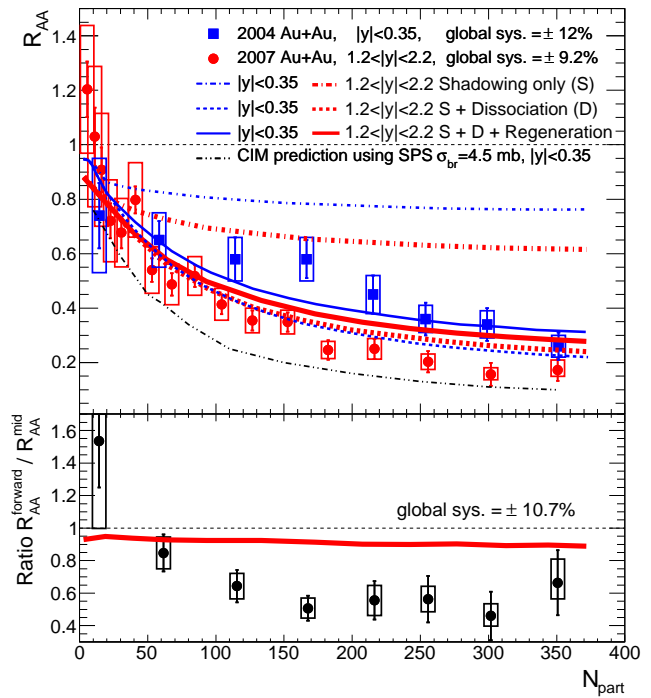}
\caption{(Color online) $\jpsi$ $\raa$ as a function of $\npart$.  Curves are
  calculations within the Co-mover Interaction Model (CIM).  The black
  dot-dot-dashed curve is the older CIM
  calculation~\cite{Capella:2005p3765} predicted from SPS data.  The
  remaining curves are from~\cite{Capella:2008p4766}, where the
  dot-dashed curve is from shadowing alone, the dashed line also
  includes dissociation in the co-moving medium, and the solid line is
  the total effect after including $\jpsi$s from regeneration.  The
  thin blue curves are calculated for midrapidity, while the
  thick red are for forward rapidity.  The lower panel contains the
  ratio of forward rapidity to midrapidity for all points and curves
  in the upper panel where both are
  calculated.}\label{fig:raa_npart_cim}
\end{figure}

\begin{figure}
\centering
\includegraphics[width=1.0\linewidth]{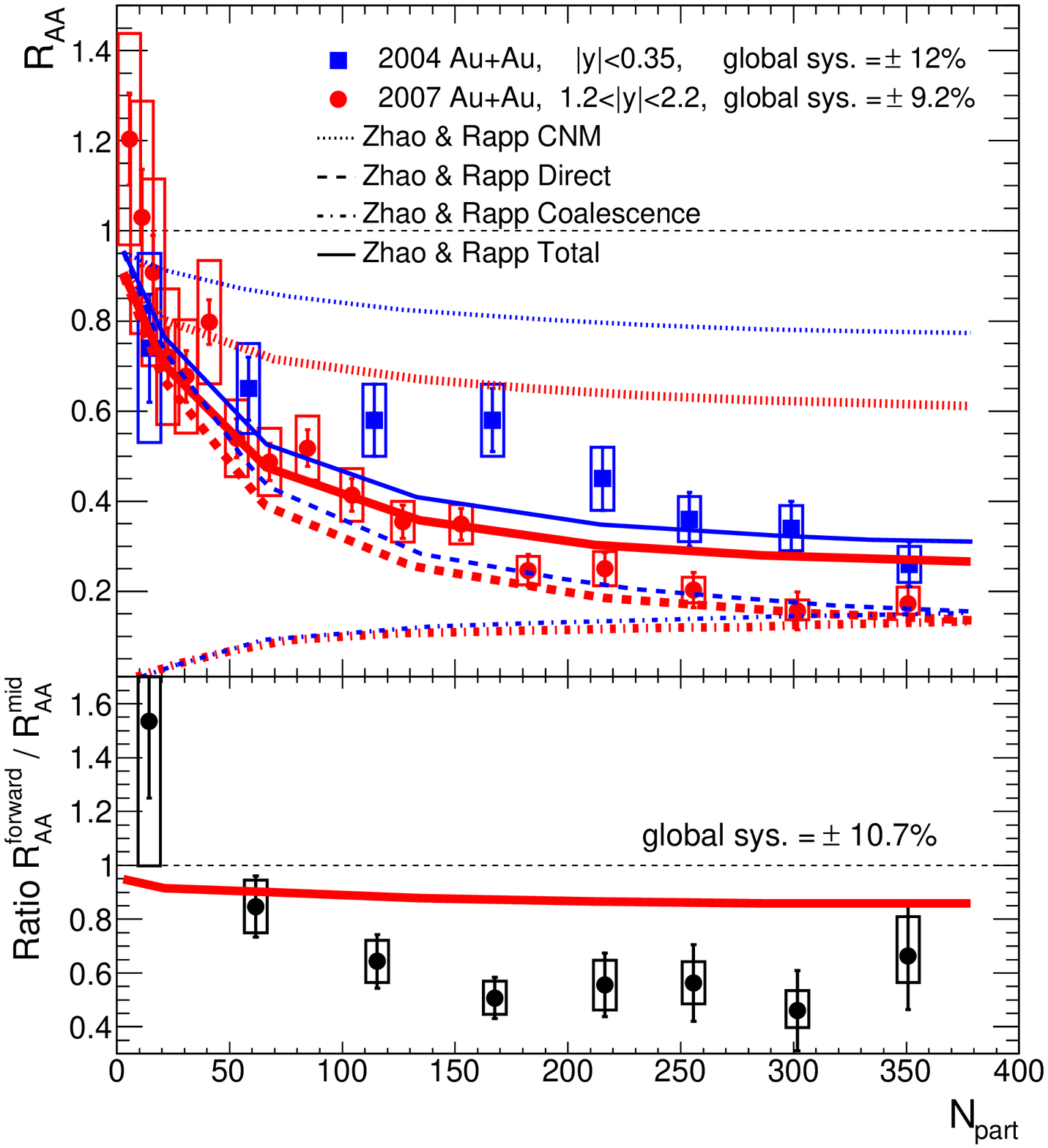}
\caption{(Color online) $\jpsi$ $\raa$ as a function of $\npart$.  Model calculations
  by Zhao and Rapp from~\cite{Zhao:2008p631,Zhao:2009p2546} are
  included for both rapidity bins, incorporating cold and hot nuclear
  matter suppression as well as coalescence of \ccbar\ pairs.  The
  various line styles represent the different contributions to the
  total as laid out in the legend, while the two thicknesses
  represent the two rapidity ranges (thin blue is midrapidity and
  thick red is forward rapidity).  The lower panel contains the ratio
  of forward rapidity to midrapidity for all points and curves in the
  upper panel.}\label{fig:raa_npart_zhao}
\end{figure}

\subsection{Comover Interaction Model}

The second calculation comes from the Comover Interaction Model 
(CIM)~\cite{Capella:2005p3765, Capella:2008p4766}.  This calculation 
uses a rate equation that accounts for $\jpsi$ breakup due to 
interactions with a dense co-moving final-state medium.  Additionally, 
the contribution from interactions with the outgoing nuclei is 
included.  No assumption is made about the nature of the co-moving 
medium, i.e.~whether it is partonic or hadronic, only that it can be 
represented by a comover density and comover-$\jpsi$ cross section 
$\sigma_{\rm co}$, for which a value of 0.65 mb was found to match the 
NA50 data and then used for the projection to \auau\ at RHIC.  The 
separate nuclear breakup cross section was taken to be 
$\sigma_{\rm br}$ = 4.5 mb. This value was taken from measurements in \pa\ 
collisions at $\sqrts$ = 27.4/29.1 GeV at the CERN-SPS. Under the 
assumption that $\sigma_{\rm br}$ is energy-independent, the value from 
those measurements was used until recently as the cold nuclear matter 
reference for heavy ion collisions at $\sqrts$ = 17.2 
GeV~\cite{Abreu:2000p3069}, as well as in~\cite{Capella:2005p3765} as the 
reference for $\sqrts$ = 200 GeV at RHIC.

However, the effective breakup cross section has now been shown to 
decrease significantly with collision energy~\cite{Lourenco:2008sk}. In 
fact, a recent measurement in \pa\ collisions at $\sqrts$ = 17.2 
GeV~\cite{Arnaldi:2009ph} yielded a value of $\sigma_{\rm br}$ = $7.6 
\pm 0.7 \pm 0.6$ mb with no anti-shadowing correction, and $\sigma_{\rm 
br}$ = $9.3 \pm 0.7 \pm 0.7$ mb with anti-shadowing corrected for.  
The value measured in \dau\ collisions at RHIC~\cite{Adare:2007gn} is 
$2.8^{+1.7}_{-1.4}$ mb (after shadowing is accounted for using EKS98 
nuclear PDFs), somewhat smaller than the 4.5 mb used 
in~\cite{Capella:2005p3765}.  The calculation, shown in 
Fig.~\ref{fig:raa_npart_cim} as the black, dot-dot-dashed curve, 
significantly overestimated the suppression measured at midrapidity for 
\auau\ collisions at $\sqrts$ = 200 
GeV~\cite{PPG068,PereiraDaCosta:2005xz}.  The suppression is stronger 
than the SPS case mainly due to the larger comover density calculated 
for RHIC.

An updated calculation~\cite{Capella:2008p4766} was then released that 
replaced the constant nuclear breakup cross section with a 
Bjorken-$x$-dependent function that goes to $\sigma_{\rm br}$ = 0 mb at 
$y$ = 0, while the same $\sigma_{\rm co}$ = 0.65 mb was used for the 
comover interactions as before. Additionally, a $\jpsi$ regeneration 
component was added that is normalized to the ratio of open charm 
production squared to $\jpsi$ production in $\pp$ collisions.  These 
new results are included in Fig.~\ref{fig:raa_npart_cim}.  The 
suppression from initial-state effects alone is much weaker at 
midrapidity than the previous calculation, due to both the change in 
the nuclear absorption, as well as an updated parametrization of 
shadowing effects.

The CNM effects (i.e.~shadowing and nuclear absorption) are much 
stronger at forward rapidity than midrapidity, due in part to the 
assertion that nuclear absorption is negligible at midrapidity.  On the 
other hand, the effects of comover dissociation and regeneration are 
stronger at midrapidity.  The combination of these three effects leads 
to predictions which are overall very similar at forward and 
midrapidity (as seen in the lower panel).

\subsection{QGP/Hadron Gas Model}

The third model we compare with is from Zhao and 
Rapp~\cite{Zhao:2008p631,Zhao:2009p2546}, which incorporates both a 
quark-gluon plasma (QGP) phase and a hadronic gas (HG) phase.  In this 
calculation, they include two different models for cold nuclear matter 
effects.  In the first case, nuclear absorption is calculated in the 
usual Glauber formalism, shadowing plus anti-shadowing are assumed to 
roughly cancel, such that the overall shadowing effects are 
encapsulated in the breakup cross section $\sigma_{\rm br}$, and $\pt$ 
broadening is included via Gaussian smearing. In the second case, the 
cold nuclear matter effects are treated as in~\cite{Capella:2008p4766}, 
supplemented with the same $\pt$ broadening model as the first case.

The thermal dissociation is modeled via a Boltzmann transport equation for 
both QGP and HG phases. The QGP is assumed to be an isentropically 
expanding cylindrical fireball.  $\jpsi$-medium interactions are assumed 
to stop at a freeze-out temperature of 120 MeV.  The final $\jpsi$ 
$\pt$ distribution is calculated by spatially integrating the final 
phase-space distribution.  The regeneration component assumes that the 
\ccbar\ is thermally equilibrated with the medium when it coalesces into a 
$\jpsi$.  Consequently, and the $\psi$ $\pt$ distribution is governed by a 
blastwave equation for the transverse flow velocity.  The normalization of 
this component is performed by plugging the initial charm densities into a 
rate equation with both gain and loss terms, and solving at the freeze-out 
time.

\begin{figure*}
\centering
\includegraphics[width=0.48\linewidth]{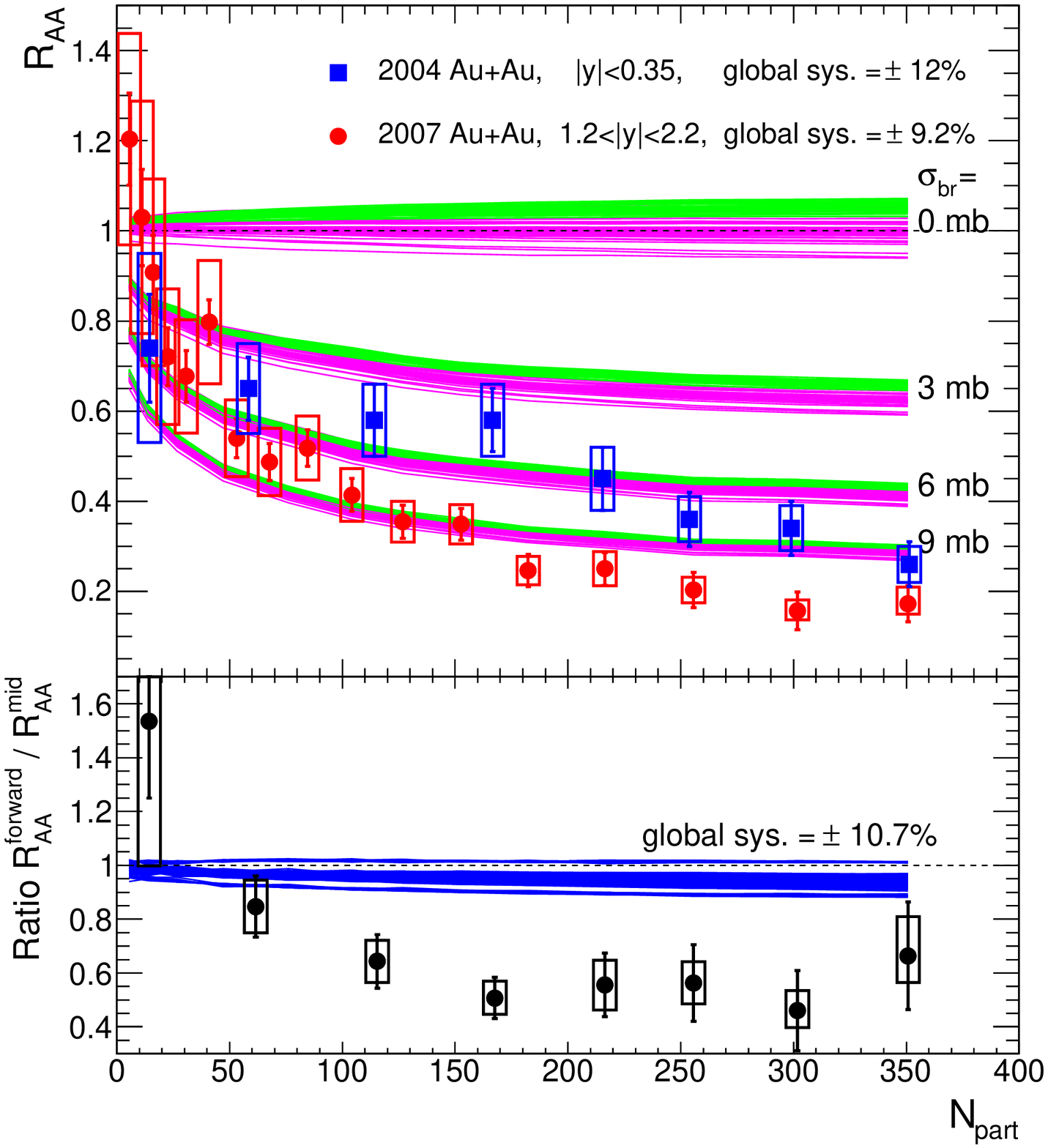}
\includegraphics[width=0.48\linewidth]{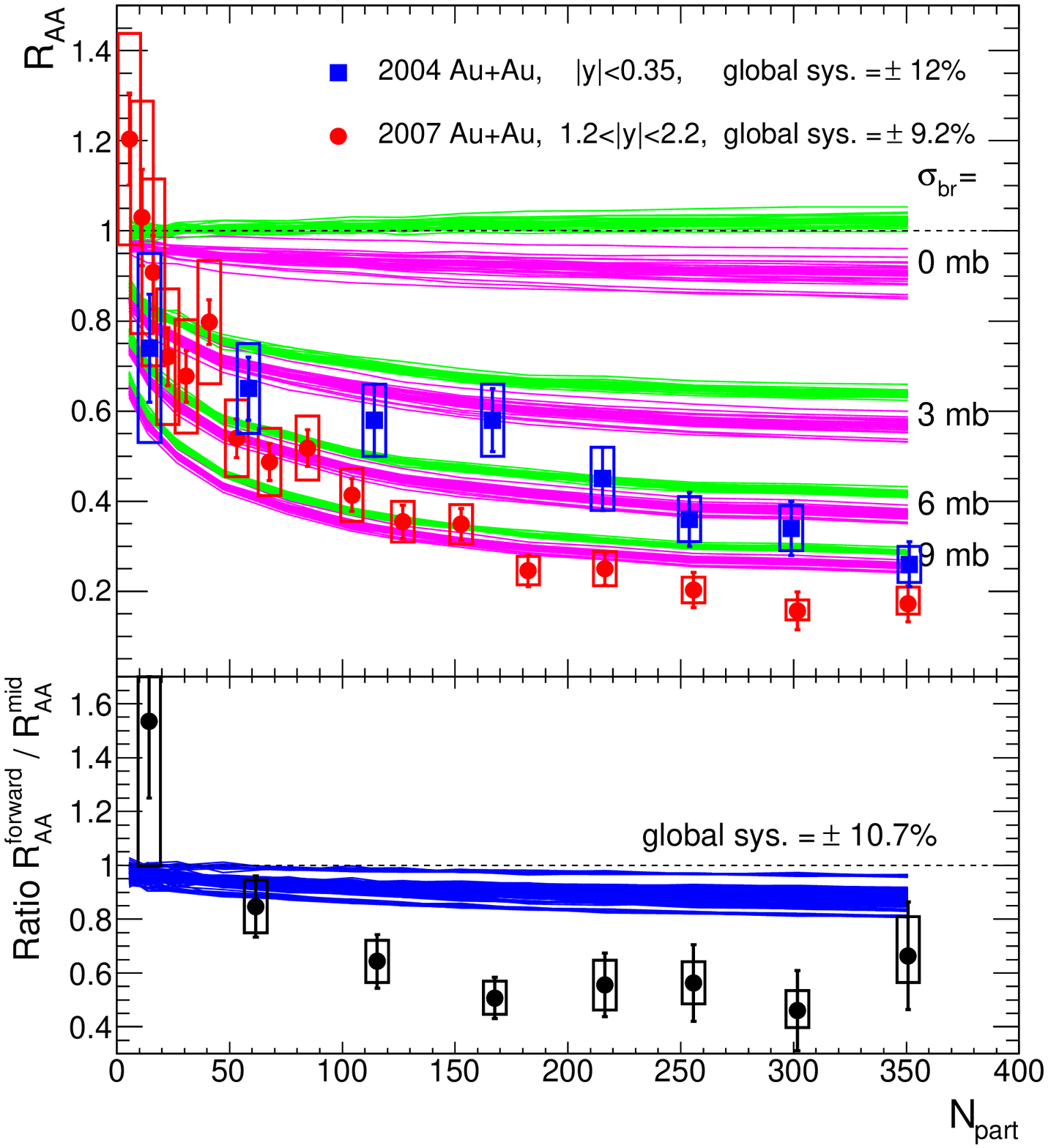}
\caption{(Color online) $\jpsi$ $\raa$ as a function of $\npart$.  
Shown in the left panel is a comparison with a calculation including 
cold nuclear matter effects (nPDF and $\sigma_{breakup}$) following the 
prescription in~\cite{Adare:2010fn,Nagle:2010}.  The green (magenta) 
lines (light gray (dark gray) in B\&W) are the result at midrapidity 
(forward rapidity) for all 31 EPS09 nPDF variations and the labeled 
value for $\sigma_{breakup}$ for 0, 3, 6, and 9 mb.  In the lower 
panel, one sees all 31 EPS09 variations times the various 
$\sigma_{breakup}$ values resulting in less than a 10\% difference 
between forward and midrapidity modification.  The right panel shows 
the same calculation, but now including initial state parton energy 
loss with a quadratic length dependence.  The chosen strength of the 
energy loss is that which most closely matches the \dau\ data as 
detailed in paper~\cite{Nagle:2010}.  The lower panels in both cases 
contain the ratios of forward rapidity to midrapidity for all points 
and curves in the upper panels.}\label{fig:raa_npart_nagle}

\end{figure*}

The calculation results, using the second case for the cold nuclear 
matter effects, are shown in Fig.~\ref{fig:raa_npart_zhao}, along 
with the separate dissociation and regeneration components.  Though the 
second cold nuclear matter case increases the suppression for more 
central events compared to the first scenario, the difference in the 
overall suppression between the two scenarios is small.  The 
qualitative trends of the calculation agree with the experimental data; 
however, the calculated suppression is very similar between forward and 
midrapidity, which is in disagreement with the data.

It is noteworthy that the regeneration component is only slightly 
larger at midrapidity in this model than at forward rapidity.  This is 
in contrast to other regeneration or recombination calculations that 
result in a significant narrowing of the $\jpsi$ rapidity distribution 
in central \auau\ events (see for 
example~\cite{Thews:2006p3770,Zhou:2009vz}).  In simple calculations, 
the $\jpsi$ recombination contribution scales as the square of the 
local charm density ($(dN_{c\overline{c}}/dy)^{2}$) and thus there is 
substantially less recombination at forward rapidity.  This modeling 
also leads to predictions of significantly larger recombination 
enhancements at the LHC where charm production is much larger. However, 
in this calculation~\cite{Zhao:2009p2546} with a full space-momentum 
distribution of charm pairs, the probability of a charm quark from one 
$c\overline{c}$ pair recombining with an anti-charm quark from another 
$c\overline{c}$ pair is suppressed because they are typically spatially 
separated which is then maintained through collective flow.  Thus, 
their recombination is dominated by the case where a $c\overline{c}$ is 
produced as a pair that would normally not form a $\jpsi$, but due to 
scattering in the medium have a re-interaction and recombine.  In this 
case, the regeneration contribution has a rapidity dependence similar 
to that of the directly produced $\jpsi$.

\subsection{Shadowing/Nuclear Absorption/Initial-state Energy Loss Model}

In addition to the above three models, we use a framework for calculating 
just the cold nuclear matter effects and extrapolating them to \auau\ 
collisions.  We begin with the prescription in~\cite{Nagle:2010} for \dau\ 
collisions, which combinines effects of nuclear-modified parton 
distributions functions (nPDFs) using the EPS09 
parameterization~\cite{Eskola:2009uj} with a rapidity-independent 
$\jpsi$-nucleon breakup cross section $\sigma_{\rm br}$, along with the 
possibility of initial-state parton energy loss.  We have extended these 
calculations to the \auau\ case using the identical code.  First, we 
include the variations of the EPS09 nPDF sets and the breakup cross 
section $\sigma_{\rm br}$.  Shown in the left panel of 
Fig.~\ref{fig:raa_npart_nagle}, is the projected cold nuclear matter 
effect from these two contributions.  The top green band shows the results 
for $\jpsi$ at midrapidity from all 31 EPS09 nPDF variations with a 
$\sigma_{\rm br} = 0$ mb.  The lower green bands are in steps of 
$\sigma_{\rm br} = 3,6,9$ mb.  The magenta bands show results for $\jpsi$ 
at forward rapidity.  The centrality dependence is not reproduced for any 
value of $\sigma_{\rm br}$ at either rapidity, though a value of 
$\sigma_{\rm br}$ greater than 6 mb (9 mb) is required to approach the 
data at mid (forward) rapidity.  In the lower panel we show all 31 EPS09 
nPDFs times 4 $\sigma_{\rm br}$ values spanning the range 
$\sigma_{\rm br}$ = 0-9 mb for the ratio of the forward to midrapidity 
suppression.  No combination of these two effects reproduces the 
modification in the rapidity shape for mid to central \auau\ collisions.  
One reason for the modest rapidity dependence is that at forward rapidity 
the $\jpsi$ production results from one low-$x$ gluon (in the nPDF 
shadowing regime) and one high-$x$ gluon (in the nPDF anti-shadowing 
regime) and the two effects largely cancel.

As discussed in~\cite{Nagle:2010}, one can attempt to improve the cold 
nuclear matter calculation agreement with the \dau\ data by including a 
parameterization of initial-state parton energy loss.  In the right 
panel of Fig.~\ref{fig:raa_npart_nagle}, we again plot all 31 EPS09 
nPDF parameterizations, 4 $\sigma_{\rm br}$ values and the quadratic 
length dependent initial-state parton energy loss that best matched the 
\dau\ data.  The initial-state parton energy loss has a minimal effect 
over this rapidity range, which is not unexpected since the effect only 
becomes significant in \dau\ for rapidity $y > 1.8$.

We note that the cold nuclear matter calculation does not
give a full description of the \dau\ data, adding some
uncertainty to its use in \auau\ collisions.  Nevertheless,
it is informative that the calculation clearly fails to
simultaneously explain the \auau\ data at forward and
midrapidity.  We observe $J/\psi$ suppression beyond that
expected from the cold nuclear matter effects included
in this calculation with the choice of a reasonable value
of 3.0-3.5 mb for $\sigma_{\rm br}$ at RHIC. 

\begin{figure}
\centering
\includegraphics[width=1.0\linewidth]{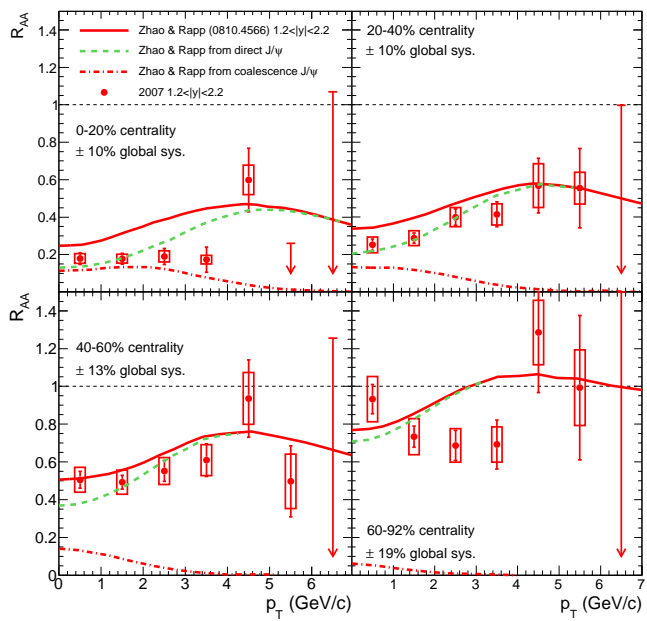}
\caption{(Color online) $\jpsi$ $\raa$ as a function of $\pt$ in four centrality
  bins.  Model calculations by Zhao and Rapp
  from~\cite{Zhao:2008p631,Zhao:2009p2546} are included for the
  forward rapidity bin, incorporating cold and hot nuclear matter
  suppression as well as coalescence of \ccbar\
  pairs.}\label{fig:raa_pt_zhao}
\end{figure}

\begin{figure}
\centering
\includegraphics[width=1.0\linewidth]{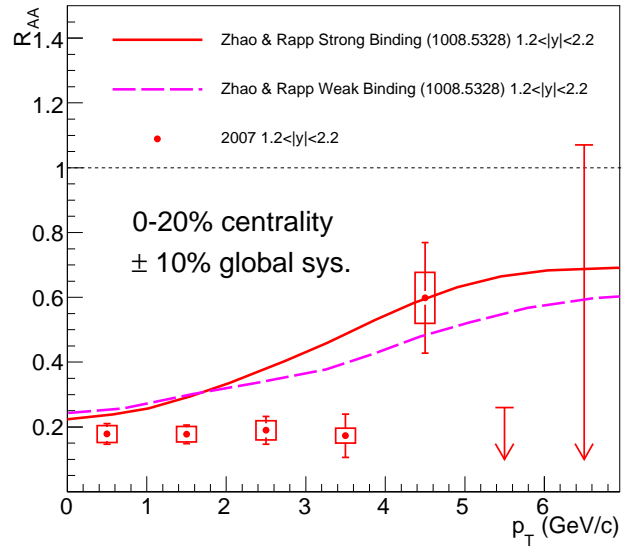}
\caption{(Color online) $\jpsi$ $\raa$ as a function of $\pt$ in the 0-20\%
  centrality bin at forward rapidity.  Overlaid are a more recent
  calculation of Zhao and
  Rapp~\cite{Zhao:2010p5227,Zhao:2010Private}.}\label{fig:raa_pt_zhao2}
\end{figure}

\subsection{$\pt$ Dependence of the Suppression}

Most of the above calculations do not include predictions for the 
$\jpsi$ suppression as a function of transverse momentum.  However, the 
calculation of Zhao and Rapp~\cite{Zhao:2009p2546} provides nuclear 
modification factors at both mid- and forward rapidity as a function of 
$\pt$.  Shown in Fig.~\ref{fig:raa_pt_zhao} are the results compared 
with our experimental data.  Their calculations indicate a moderate 
rise in $\raa$ versus $\pt$ predominantly due to the Cronin 
effect~\cite{Antreasyan:1979}. In fact, in other recombination models 
the enhancement is limited to low $\pt$~\cite{Thews:2006p3770}, and in 
this calculation the recombination contribution drops off beyond $\pt 
\gtrsim 3$ GeV/$c$. At low $\pt$ in the most central bin, the 
suppression from this calculation is too weak by up to a factor of two.

More recently, Zhao and Rapp~\cite{Zhao:2010p5227,Zhao:2010Private} 
have modified this calculation to include feed-down from $B$-mesons and 
a reduced suppression at higher $\pt$ due to the longer formation time 
of the preresonance state to the $\jpsi$ from time dilation. These 
contributions serve to increase $\raa$ at higher $\pt$ compared to the 
previous calculation, and are compared to the forward rapidity data in 
Fig.~\ref{fig:raa_pt_zhao2}.  The current lack of statistics in the 
data at $\pt > 5$ GeV/$c$ precludes a confirmation of this effect.

It should also be noted that in the new calculation by Zhao and Rapp, 
cold nuclear matter effects are handled differently than in the earlier 
calculation. An effective absorption cross section of 3.5 (5.5) mb at 
$y$ = 0 (1.7) is used to account for the combined effects of shadowing 
and breakup. These effective cross sections are obtained from 
comparison with recent PHENIX \dau\ data. They argue that the larger 
effective breakup cross section at forward rapidity is most likely 
associated with shadowing effects, and so reflects a suppression of the 
number of charm pairs relative to midrapidity.  Therefore, the 
additional effective absorption at forward rapidity is associated with 
a reduction in the open charm yield as well as the $\jpsi$ yield, thus 
also reducing the $\jpsi$ regeneration contribution. As a result of the 
ad hoc use of a larger effective absorption cross section at forward 
rapidity, this new calculation produces a forward to midrapidity $\raa$ 
ratio of about 0.7 in central collisions, which is in better agreement 
with the data.  Again, this is entirely due to the cold nuclear matter 
effects, and in fact the hot nuclear matter suppression in this 
calculation is almost identical between forward and midrapidity.
 
A second model of interest to the $\pt$ dependence of $\jpsi$ 
production is the so-called Hot Wind model~\cite{Liu:2007p625}.  This 
model predicts a decrease in $\jpsi$ $\raa$ at higher $\pt$ in 
semi-central events, based on a modification of the screening length 
due to the relative velocity between the $\jpsi$ and the medium. 
However, there is no quantitative calculation available at forward 
rapidities, and there is no evidence for such an effect in the $\pt$ 
range covered by the present data.

\section{Summary and Conclusions}

We have presented new and more precise measurements of $\jpsi$ nuclear 
modification at forward rapidity in \auau\ collisions at $\sqrts = 200$ 
GeV.  The results confirm our earlier published findings of a larger 
suppression at forward compared with midrapidity.  This, combined with 
the similar suppression of $\jpsi$ at midrapidity between RHIC and 
lower energy measurements, remains an outstanding puzzle in terms of a 
full theoretical description.

Due to the lack of a comprehensive and consistent understanding of the 
numerous cold nuclear matter effects and their extrapolation to 
nucleus-nucleus collisions, extracting a quantitative measurement of 
the hot nuclear matter effects is not possible at the present time. 
However, it is clear that the observed suppression is attributable to 
hot nuclear matter effects, as the suppression in \auau\ collisions is 
larger than predicted by the current models of CNM effects.
It will be useful for all calculations to include the transverse 
momentum dependence for future comparisons, especially as the 
experimental uncertainties, particularly at high $\pt$, will only 
improve in the future.



\end{document}